\documentclass[aps,prd,preprint,groupedaddress]{revtex4-1}
\usepackage{graphicx}
\usepackage{amssymb}
\usepackage{amsmath}
\usepackage{epstopdf}

\newcommand{\om}{\omega}

\newcommand{\be}{\begin{equation}}
\newcommand{\ee}{\end{equation}}
\newcommand{\bea}{\begin{eqnarray}}
\newcommand{\eea}{\end{eqnarray}}
\newcommand{\eps}{\epsilon}

\newcommand{\eg}{{\em e.g. }}
\newcommand{\ie}{{\em i.e. }}

\begin{document}


\title{Precision tests of General Relativity with Matter Waves}


\author{Michael A. Hohensee}
\email[]{hohensee@berkeley.edu}
\affiliation{Department of Physics, 366 Le Conte Hall MS 7300, University of California, Berkeley, California 94720, USA}
\author{Holger M\"uller}
\affiliation{Department of Physics, 366 Le Conte Hall MS 7300, University of California, Berkeley, California 94720, USA}


\date{\today}

\begin{abstract}
We review the physics of atoms and clocks in weakly curved spacetime, and how each may be used to test the Einstein Equivalence Principle (EEP) in the context of the minimal Standard Model Extension (mSME).  We find that conventional clocks and matter-wave interferometers are sensitive to the same kinds of EEP-violating physics.  We show that the analogy between matter-waves and clocks remains true for systems beyond the semiclassical limit.  We quantitatively compare the experimentally observable signals for EEP violation in matter-wave experiments.  We find that comparisons of $^{6}$Li and $^{7}$Li are particularly sensitive to such anomalies.  Tests involving unstable isotopes, for which matter-wave interferometers are well suited, may further improve the sensitivity of EEP tests.
\end{abstract}


\maketitle


\section{Introduction}

The gravitational redshift is an important prediction of general relativity, was the first experimental signature considered by Einstein in 1911~\cite{Einstein1911}, and its experimental verification remains central to our confidence in the theory.  Clock comparison tests~\cite{Hafele:1972,Chou} have reached accuracies of $7$ parts in $10^{5}$~\cite{Vessot}, while experiments based on matter waves, in which a redshift anomaly would modify the Compton frequency of material particles, have reached an accuracy of $7$ parts in $10^{9}$~\cite{redshift,Poli,redshiftPRL}.  These experiments complement a wide array of other tests of the equivalence principle, including tests of the universality of free fall (UFF) and local Lorentz invariance~\cite{Schlamminger:2008, datatables}.   We briefly review the physics of the gravitational redshift and the acceleration of free fall relevant to clocks, and moving test masses, both quantum and classical, in the limit of a weak, static gravitational potential.  Using the mSME~\cite{KosteleckyGravity,KosteleckyTassonPRL,KosteleckyTasson},
we determine the phenomenological parameters for EEP violation that are constrained by redshift and UFF tests.  Focusing on those terms of the mSME that are only observable by gravitational experiments, we find that using metastable nuclides in a matter-wave interferometer may offer improved sensitivity to such effects.

\section{Action and the Gravitational Redshift}

If $x^{\mu}(\lambda)$ are the coordinates of a clock moving on a path parameterized by the affine parameter $\lambda$ through space-time, and $g_{\mu\nu}$ is the metric, the proper time $d\tau$ experienced by a locally inertial clock as it moves a distance $dx^{\mu}$ is \cite{MTW}
$d\tau=\sqrt{-g_{\mu\nu} dx^{\mu}dx^{\nu}}/c$. If the metric differs between two points $x^{\mu}_{1,2}$ in spacetime, then two otherwise identical oscillators with the same proper frequency $f_{0}$ can appear to an observer to tick at different frequencies $f_{1,2}$, since the relationship between the proper time $\tau$ and coordinate time $t=x^{0}/c$ is a function of position.  The difference frequency $\delta f=f_{1}-f_{2}$ for locally inertial clocks moving with nonrelativistic velocities $\vec{v}_{1}$ and $\vec{v}_{2}$ in a weak static gravitational potential $\phi_{i}=-MG/|\vec{r}_{i}|$ becomes
\begin{equation}
\frac{\delta f}{f_{0}}=\frac{\phi_{1}-\phi_{2}}{c^{2}}-\frac{v_{1}^{2}-v_{2}^{2}}{2c^{2}}+O\left(c^{-3}\right).\label{eq:moveclock}
\end{equation}
The first term is the gravitational redshift, first measured by Pound and Rebka in 1960~\cite{PoundRebka}. The second is the time dilation due to the clock's motion, and can be subtracted if the trajectories are known.  This equation is universal up to terms proportional to $c^{-2}$; at $O(c^{-3})$ and beyond, measurements of the instantaneous $\delta f$ depend upon how the signals carrying the clocks' frequencies propagate and where the comparison takes place.

It has recently been argued~\cite{comment} that matter wave experiments do not constitute tests of the gravitational redshift, but should rather be understood as probes of UFF.  We note that similar arguments have been leveled at clock comparison tests in the past~\cite{schiff}, and that tests of UFF and the gravitational redshift are generally not independent of one another in any theory which conserves energy and momentum~\cite{nordtvedt}. In order to explain the analogy between clocks and matter waves, let us consider two clocks which are initially synchronized to have identical phase $\varphi_0=0$ at $t=0$ and then transported to the same point along different paths, where they are compared. Then, the phase of clock $1$ relative to clock $2$ is $\delta \varphi_f\equiv\varphi_{1}-\varphi_{2}= \int 2\pi\, \delta f\,dt$, is given by
\begin{equation}
\delta\varphi_f=\om_0\int_{0}^{T}dt\,\left(\frac{\vec r_{12} \cdot \vec g}{c^{2}}-\frac{v_{1}^{2}-v_{2}^{2}}{2c^{2}}\right),\label{eq:freeevrocket}
\end{equation}
where $\om_0=2\pi f_0$ and $\vec r_{12}=\vec r_1-\vec r_2$; we have specialized to a homogenous gravitational field so that $\phi_{1}-\phi_{2}=\vec g\cdot \vec r_{12}$. If the clocks are freely falling, then their motion is an extremum of their respective actions, given by~\cite{MTW}
\begin{equation}
S_{i}=\int m c^{2}\, d\tau\approx\int mc^{2}\left[1+\frac{\phi_{i}}{c^{2}}-\frac{v_{i}^{2}}{2c^{2}}\right] dt,\label{eq:freeaction}
\end{equation}
where, since we work in the non-relativistic limit, we take the clocks' coordinate time to be $t(\lambda)=\lambda$, and thus use $dt$ in the place of $d\lambda$.

Gravity also acts upon the quantum phase of matter waves~\cite{Colella, Bonse, Horne, Werner, Littrell}.  In Feynman's path integral formulation~\cite{FeynmanHibbs} of quantum mechanics, the wavefunction $\psi$ for a particle with mass $m$ at $t_{B}$ is obtained from its value at $t_{A}$ according to
\bea
\psi(t_B,\vec x_B)&=&\int d^{3}x K(t_{B},\vec x_{B};t_{A},\vec x_{A})\psi(t_{A},\vec{x}_{A}), \nonumber \\
K(t_{B},\vec x_{B};t_{A},\vec x_{A})&=&\int_{(t_{A},\vec x_{A})}^{(t_{B},\vec x_{B})}\exp\left[(i/\hbar)\int_{\tau_{A}}^{\tau_{B}} mc^2 d\tau\right]\mathcal{D}\vec{x}(\tau),
\eea
where $\mathcal D\vec{x}(\tau)$ indicates that the integral is taken over all paths.  
In the semiclassical limit, the matter-wavepacket $\psi$ follows the classical path of least action, and acquires a phase shift $e^{i\varphi}$, with
$\varphi=\frac{S}{\hbar}=\omega_{C}\int_{A}^{B} d\tau$,
where $\omega_{C}=\frac{mc^{2}}{\hbar}$.  We thus conclude that the relative phase accumulated by two identical matter-wavepackets that travel along separated paths is the same, up to a constant factor of $\omega_{C}/\omega_{0}$, as that acquired by two conventional clocks which follow the same trajectory.  Note that although this expression applies to the semiclassical limit, we need not work in this limit to interpret matter-wave experiments as redshift tests.  In the appendix, we derive the non-relativistic Schr\"odinger equation from the path integral in the weak gravitational field limit.

For any pair of conventional clocks, (\eg electronic oscillators referenced to a microwave, or optical transition) the phase difference accumulated over a given period of coordinate time is a small fraction of the total quantum phase $(S_{1}-S_{2})/\hbar$ they may accumulate.  Since the phase of a matter-wave oscillates at the Compton frequency ($\sim 10^{24}$ Hz), the intrinsic sensitivity of a matter-wave interferometer to variations in the proper time is between $10^{10}$ and $10^{14}$ times greater than that of a conventional clock.  The greater precision in the phase readout and the greater separation available to optical and microwave clocks can bridge part, but not all of this divide.

\section{QED and the Gravitational Redshift\label{sec:photon}}

With the exception of the Hafele-Keating experiment~\cite{Hafele:1972}, all redshift tests prior to the advent of matter-wave interferometry have used electromagnetic signals to compare the relative ticking rate of two clocks.  Since a Mach-Zehnder matter-wave interferometer more closely resembles the Hafele-Keating experiment in that the relative clock rates are never encoded in the frequency of a photon, one might reasonably be concerned that matter-wave interferometers might be unable to observe anomalous redshift physics detectable by more conventional tests.  In the absence of an anomalous gravitational coupling to spin (\ie torsion), or wavelength-dependent gravitational coupling to light, and in the limit that the photon remains massless in the vacuum, this concern can be resolved by applying general covariance, \ie our freedom to choose our coordinate system.

The properties of a curved spacetime metric and of a lone field propagating within that metric are never directly observable~\cite{Mattingly:2005,KosteleckyTasson}.  Instead, we must infer these properties by comparing the effects of the metric on several different fields.  General covariance affords us complete freedom to choose the coordinate chart upon which the metric tensor $g_{\mu\nu}$ is defined, and the freedom to arbitrarily define the coordinates of the local Lorentz frame at a single point in spacetime.  Any anomaly in the coupling of light to gravity can be expressed as a modification of the metric tensor $g_{\mu\nu}$.  This modification can be formally eliminated from the electromagnetic sector of the theory by a redefinition of the local Lorentz frame, so that photons behave conventionally. This moves the anomalous QED couplings into the physics of all other particle fields.  The existence of any photon-mass and spin-dependent anomalies, while not considered in detail here, has been strongly constrained by spectro-polarimetric studies of light emitted by distant gamma ray bursts~\cite{Kostelecky:2006}.  More recently, a broader class of wavelength-dependent QED anomalies has also been tightly bounded by astrophysical observations~\cite{Kostelecky:2008a}.  While neither study explicitly considered anomalies arising from gravitational interactions, their results suggest that such effects, if they exist, are likely to be extremely small in any terrestrial experiment.

\section{Equivalence Principle Tests in the Standard Model Extension}

In the non-relativistic limit, the motion and gravitational redshift experienced by a freely falling particle are determined by the same element $g_{00}$ of the metric tensor.  It is therefore no surprise that tests of UFF and the gravitational redshift are not independent of one another.  Indeed, the two must be linked in any energy-conserving theory~\cite{nordtvedt}.  We will now explore this relationship in the context of the minimal gravitational standard model extension~\cite{KosteleckyGravity,KosteleckyTassonPRL,KosteleckyTasson}.  The EEP requires that the laws of physics be the same in all local inertial frames, no matter where they are or how fast they are moving, and that gravity must act through the curvature of spacetime alone, affecting all particles in the same way~\cite{MTW,TEGP}.  Both clock comparison and matter-wave interferometer tests can be used to test the EEP, and their results can be used to quantitatively restrict the degree to which weak position- or velocity-dependent effects described by the mSME are consistent with the observed laws of physics.  The mSME framework is formulated from the Lagrangian of the standard model by adding all Lorentz- and CPT violating (and thus EEP-violating) terms that can be formed from the standard model fields and Lorentz tensors~\cite{ColladayKostelecky}.  Some of these terms, which can represent the vacuum expectation values of heretofore unknown fields, are only detectable via gravitationally-induced fluctuations in their mean values~\cite{KosteleckyTasson}.  They can also contribute to the metric tensor $g_{\mu\nu}$ via their effect on the stress-energy tensor.  Since the effective particle Lagrangian that results is not an explicit function of space or time, the mSME conserves energy and momentum.

Most, but not all, coefficients in the gravitational mSME produce Lorentz violation that is measurable in flat space-time.  We focus on an isotropic subset of the theory and thereby upon some of the comparatively weakly constrained flat-space-observable terms, and the dominant elements of other EEP-violating vectors that are hard to detect with non-gravitational tests.  Up to $O(c^{-2})$, isotropic spin-independent EEP violation is governed by the 
six coefficients $\alpha(\bar a^w_{\rm eff})_0$ and $(\bar c^w)_{00}$, where the superscript $w$ may take the values $e,n,p$, indicating that the new physics enters via the action of the electron, neutron, or proton fields, respectively.  As the subscripts suggest, these respective coefficients are elements of a four-vector and a four-tensor: the other elements of which would give rise to spatially anisotropic anomalies.  These coefficients generate measurable violations of EEP in two ways: First, they modify the effective value of $g_{00}$ for the electrons, neutrons, and protons which make up a clock or moving test particle.  This channel is responsible for most of the signal in experiments which measure the total phase accumulated by a test particle's wavefunction, or which compare the effective gravitational acceleration of different objects.  It also contributes to the signal in conventional clock comparison tests by perturbing the motion of any test mass used to map the 
gravitational potential $\phi$.  These terms can also modify the rest-frame energy and energy levels of composite systems as a function of the gravitational potential, shifting the Compton and transition frequencies of a bound system in a species and state-dependent manner.  This is the primary signal available to EEP tests using conventional clocks.  These position-dependent binding energy shifts also produce correction to the motion of the freely falling composite particle. While this correction is small, it is important because it increases the difference between the linear combinations of mSME coefficients constrained by individual experiments. While the first mechanism is a simple function of the number of electrons, neutrons and protons in any given composite particle, estimates of the second mechanism for EEP-violation determined by the particle's internal structure, discussed in more detail below.

Without loss of generality, we choose coordinates such that light propagates in the usual way through curved spacetime (see Sec.~\ref{sec:photon}).  The Lorentz-violating properties of an object $T$ composed of $N^w$ of the neutrons, protons, and electrons can often be represented by effective coefficients
\be\label{effectiveac}
(\bar c^{\rm T})_{\mu\nu}=\sum_w \frac{N^w m^w}{m^{\rm T}} (\bar c^w)_{\mu\nu},\quad (\bar a_{\rm eff}^{\rm T})_\mu =\sum_w N^w (\bar a_{\rm eff}^w)_\mu,
\ee
where we have neglected Lorentz-violating contributions from particles other than photons mediating the binding forces (\emph{e.g.} W-bosons, $\pi$-mesons, etc.).  These Lorentz vectors and tensors are defined in one particular inertial reference frame.  Although it is conventional to adopt a sun-centered celestial equatorial reference frame \cite{BaileyKostelecky} when performing such analyses, the distinction between it and any Earth-centered frame is unimportant to a derivation of the effects of the isotropic subset of the minimal gravitational SME up to terms appearing at higher powers of $1/c$, and will not be made here.

As derived in~\cite{KosteleckyTasson}, the effects of Lorentz symmetry violation on the motion of a test particle, up to Post-Newtonian order PNO(3), as defined by their suppression by no more than 3 powers of $1/c$, are described by the particle action
\begin{equation}
S = \int m^{\rm T}c\left( \sqrt{-\left(g_{\mu\nu}+2\bar c^{\rm T}_{\mu\nu}\right)\tfrac{dx^{\mu}}{d\lambda}\tfrac{dx^{\nu}}{d\lambda}}
+\frac{1}{m^{\rm T}}\left(a^{\rm T}_{\rm eff}\right)_{\mu}\tfrac{dx^{\mu}}{d\lambda}\right)d\lambda,\label{eq:smeaction}
\end{equation}
where $(a^{\rm T}_{\rm eff})_{0}= (1-2\phi \alpha)(\bar{a}^{\rm T}_{\rm eff})_{0}$, and $(a^{\rm T}_{\rm eff})_{j}=(\bar a^{\rm T}_{\rm eff})_{j}$, 
for a non-rotating spherical source with gravitational potential $\phi$.  The $\left(\bar a_{\rm eff}^{\rm T}\right)_{\mu}$ vector, where the overbar indicates the value of $(a_{\rm eff}^{\rm T})_{\mu}$ in the absence of gravity, is typically unobservable in non-gravitational experiments, as it can be eliminated from the action by a global phase shift.  If $(a^{\rm T}_{\rm eff})_{\mu}$ has a non-minimal coupling (parameterized here by $\alpha$) to the gravitational potential, however, it does not drop out of the action under such a field redefinition, and produces observable effects.  In general, $g_{\mu\nu}$ is itself modified by the contributions of the pure gravity sector coefficients and any Lorentz-symmetry violating terms in the action for the gravitational source body. We consider only experiments performed in the Earth's gravitational field, and thus neglect the effects of such modifications of $g_{\mu\nu}$ as being common to all experiments.  The isotropic subset $(\bar a^{w}_{\rm eff})_{0}$ and $(\bar c^{w})_{00}$ is of particular interest because the former can only be observed by gravitational tests, and are not yet individually constrained; while the $(\bar c^{w})_{00}$, though measurable in non-gravitational experiments, are comparatively weakly constrained relative to the $(\bar c^{w})_{0j}$ and $(\bar c^{w})_{jk}$ terms.

The expansion of Eq. (\ref{eq:smeaction}) up to PNO(2) terms, dropping the constant term associated with the rest particle mass, and redefining $m^{\rm T}\rightarrow m^{\rm T}[1+\tfrac 53 (c^{\rm T})_{00}]$, yields
\begin{equation}\label{lagrangesimplified}
S=\int m^{\rm T} c^{2}\left(\frac{\phi}{c^{2}}\left[1-\frac{2}{3}\left(c^{\rm T}\right)_{00}+\frac{2\alpha}{m^{\rm T}}\left(\bar{a}^{\rm T}_{\rm eff}\right)_{0}\right]
-\frac{v^{2}}{2c^{2}}\right)dt,
\end{equation}
where $v$ is the relative velocity of the Earth and the test particle.  Thus, at leading order, a combination of $\left(c^{\rm T}\right)_{00}$ and $\alpha\left(\bar{a}^{\rm T}_{\rm eff}\right)_{0}$ coefficients rescale the particle's gravitational mass (proportional to $\phi$) relative to its inertial mass (proportional to $v^{2}$).

\section{Experimental Observables}

The gravitational acceleration $g^{\rm T}$ of a test mass $m^{\rm T}$ is obtained by finding the extremum of Eq.~(\ref{lagrangesimplified}), so that
\begin{equation}
g^{\rm T}=g(1+\beta^{\rm T}),\quad\beta^{\rm T}\equiv \frac{2\alpha}{m^{\rm T}}\left(\bar a^{\rm T}_{\rm eff}\right)_{0}-\frac{2}{3}(\bar c^{\rm T})_{00}.\label{eq:beta}
\end{equation}
Thus the test mass moves in the gravitational potential $\phi$ as if it were actually in a rescaled potential $\phi'=(1+\beta^{\rm T})\phi$.  The $(\bar c^{w})_{00}$ terms can also give rise to position-dependent shifts in the binding energy of composite particles.  Appearing at $O(c^{-4})$ in the expansion of Eq.~(\ref{eq:smeaction}), terms proportional to $v^{2}\phi(\bar c^{w})_{00}$ produce an anomalous $\phi$-dependent rescaling of a particle's inertial mass.  Though these terms are in most cases negligible for systems of non-relativistic, gravitationally bound particles, the internal velocities of the constituents of a composite particle held together by electromagnetic or nuclear forces are large enough to make the $v^{2}\phi(\bar c^{w})_{00}$ terms significant.  To leading order, in gravitational fields that are weak compared to the non-gravitational binding forces, it has been shown~\cite{KosteleckyTasson} that the bound particles' equations of motion are unchanged save for the substitution
\begin{equation}
\frac{1}{m^{w}}\rightarrow \frac{1}{m^{w}}\left[1+3\phi+\frac{5}{3}(\bar c^{w})_{00}-\frac{13}{3}\phi(\bar c^{w})_{00}\right],\label{eq:masssub}
\end{equation} 
causing the energy (as measured in its local frame) of a bound system of particles to vary as a function of the gravitational potential $\phi$.  For a clock referenced to a transition between different bound states of a system of particles, the substitution in Eq.~(\ref{eq:masssub}) gives rise to an anomalous rescaling of its measured redshift by a factor of $1+\xi_{\rm clock}$~\cite{KosteleckyTasson, redshiftPRL}.  The $\xi_{\rm clock}$ factor may be different for clocks referenced to different transitions, depending upon how the bound system's energy levels scale with the constituent particles' masses.  The value of $\xi_{\rm clock}$ for a Bohr transition in hydrogen has been estimated to be~\cite{KosteleckyTasson}
\begin{equation}
\xi_{\rm Bohr\, H}=-\frac{2}{3}\frac{m^{p}(\bar c^{e})_{00}+m^{e}(\bar c^{p})_{00}}{m^{e}+m^{p}},
\end{equation}
since this energy is proportional to the reduced mass $m^{e}m^{p}/(m^{e}+m^{p})$.  Energy conservation and the principle of stationary action requires that this effect also contribute to the motion of the composite particle~\cite{nordtvedt}, since any increase in the energy of a given configuration of composite particle with increasing $\phi$ must be offset by an increase in the amount of work necessary to elevate it, implying that the effective $g^{\rm T}$ for the composite system is also increased. Thus the fractional modification of the effective gravitational force acting on the hydrogen atom due to corrections of the Bohr energy would be $(2/3)(\bar c^{e})_{00}R_{\rm E}/(m^{\rm H}c^{2})\sim 10^{-8}(\bar c^{e})_{00}$.  In general, this effect scales as $(m'/m^{\rm T})\xi$, where $m'/m^{\rm T}$ is the ratio of the relevant binding energy to the bound particle's rest mass.  Even for atoms with higher $Z$, this is as small as $\sim10^{-7}\xi_{\rm Bohr\, H}$.  Note that the exact value of the binding energy correction to the motion depends upon the details of the bound system.  Contributions from $\phi$-dependent variations $\xi_{\rm nuc.}^{\rm bind}$ in the binding energy of the nucleus can be substantially larger, as the mass defect of many nucleons can represent between $0.1\%$ and $1\%$ of an atom's overall mass. The form of $\xi_{\rm nuc.}^{\rm bind}$ depends on the details of the atomic nucleus, and is model dependent.

All EEP tests compare the action of gravity on one system to its effects on another.  Relative, or null redshift tests compare the frequencies of two different clocks as they are moved about in the gravitational potential, and the precision to which they agree with one another constrains the difference $\xi_{\rm clock\:1}-\xi_{\rm clock\:2}$.  Tests involving the gravitationally-determined motion of two matter-wave clocks~\cite{redshiftPRL} or test masses constrain the difference $[\beta^{\rm T}_{1}+(m'_{1}/m^{\rm T}_{1})\xi^{\rm T}_{\rm binding\:1}]-[\beta^{\rm T}_{2}+(m'_{2}/m^{\rm T}_{2})\xi^{\rm T}_{\rm binding\:2}]$, where $m'_{j}$ is the binding energy of the test particle $j$.  Clock comparison tests in which the clocks' motion is not determined by the gravitational potential ({\emph e.g.}, they are at rest, or on continuously monitored trajectories, as in Gravity Probe A~\cite{Vessot} or the proposed ACES mission~\cite{ACES}) limit the difference $\xi_{\rm clock}-[\beta^{\rm grav}+(m'_{\rm grav}/m^{\rm grav})\xi^{\rm grav}_{\rm binding}]$, where the superscript ``grav'' denotes terms applicable to the gravimeter used to measure the potential $\phi$ used to calculate the expected value of the clock's redshifted signal.  In principle, the gravimeter could also be another clock.  See~\cite{redshiftPRL} for a more detailed analysis relevant to some specific tests of EEP.

\section{Sensitivity to the $(\bar a^{w}_{\rm eff})_{0}$ coefficients}

The $(\bar a^{w}_{\rm eff})_{\mu}$ coefficients of the gravitational mSME are of particular interest because they are difficult to observe in non-gravitational experiments~\cite{ColladayKostelecky}.  In a flat spacetime, these terms can be eliminated from each particle's Lagrangian by a global phase shift.  This is not necessarily the case in a curved spacetime~\cite{KosteleckyTasson,KosteleckyTassonPRL}; gravitationally induced fluctuations (proportional to the potential $\phi$ and an arbitrary gravitational interaction constant $\alpha$) in $(a^{w}_{\rm eff})_{\mu}$ are observable.
\begin{figure}
\begin{center}
\includegraphics[width=0.8\textwidth]{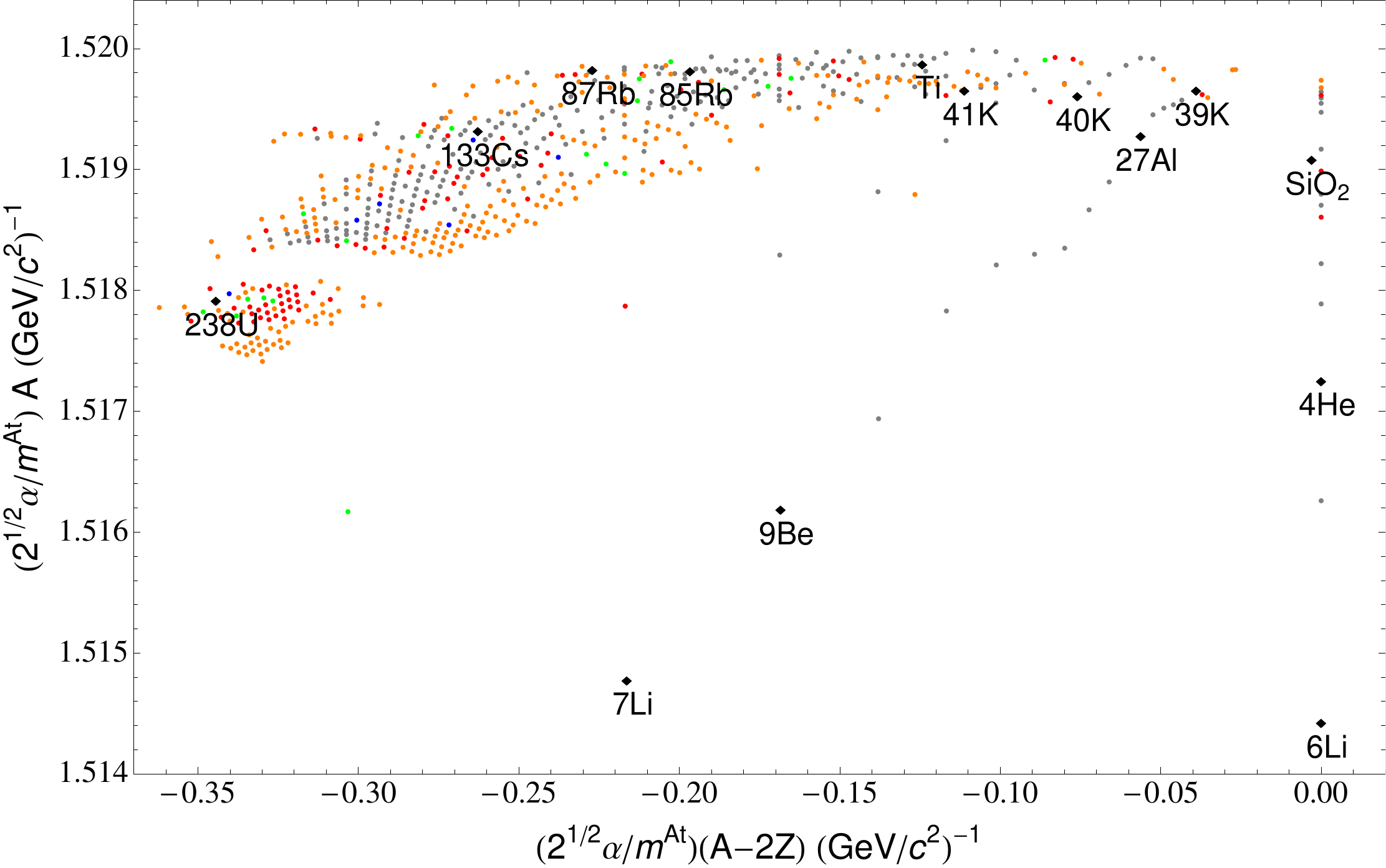}
\caption{Sensitivity to $(\bar a_{\rm eff}^{p+e+n})_{0}$ (vertical axis) and $(\bar a_{\rm eff}^{p+e-n})_{0}$ (horizontal axis) for different nuclear isotopes.  Experiments that compare two nuclides that are widely separated on this plot have greater sensitivity than those that use neighboring nuclides.  Gray points indicate stable isotopes, while blue, green, and orange points indicate isotopes with lifetimes of over 1 Gyr, 1 Myr - 1 Gyr, or 1 yr - 1 Myr, respectively.  Red points indicate isotopes with lifetimes measured in hours.  The sum and difference factors for Ti and SiO$_{2}$ are defined for objects made with natural isotopic abundances.  Not shown are the coefficients for $^{1}$H, $^{2}$H, $^{3}$H, or $^{3}$He.  Nuclide data is taken from~\cite{nubase}.    }
\label{fig:aplot}
\end{center}
\end{figure}

These coefficients are readily found in any test sensitive to the $\beta^{\rm T}$ of one or more test particles, as given by Eqs. (\ref{effectiveac}) and (\ref{eq:beta}).  Since different materials are made of different numbers of neutrons, protons, and electrons, UFF or matter-wave tests involving different species can be used to set limits on the $\alpha(\bar a^{w}_{\rm eff})_{0}$ coefficients.  Practical limitations, however, can make it difficult to set independent constraints on all three terms.  Tests involving neutral particles, for example, are only sensitive to the sum $(\bar a^{e+p}_{\rm eff})_{0}\equiv(\bar a^{p}_{\rm eff})_{0}+(\bar a^{e}_{\rm eff})_{0}$.  The fractional $(\bar a^{w}_{\rm eff})_{0}$-dependent shift in the gravitational potential reduces to
$\frac{2\alpha}{m^{\rm T}}\left((A-Z)(\bar a^{n}_{\rm eff})_{0}+Z(\bar a^{e+p}_{\rm eff})_{0}\right)$.  

Placing constraints upon these two neutral-particle parameters is further complicated by the fact that the number of neutrons $(A-Z)$ relative to the number of protons $Z$ typically satisfies $(A-Z)/m^{\rm At.}\sim (1.06 c^{2}/GeV)-Z$ for stable nuclei.  This often results in a significant suppression of the EEP-violating signal proportional to $\alpha(\bar a^{n}_{\rm eff})_{0}$ and $\alpha(\bar a^{e+p}_{\rm eff})_{0}$ when the effect of gravity on different systems is compared.  It is therefore useful to consider which combination of atomic species might best be employed to obtain limits on the $\alpha(\bar a^{w}_{\rm eff})_{0}$ coefficients.  Most experiments are primarily sensitive to the combination $(\bar a_{\rm eff}^{p+e-n})_{0}\equiv\alpha(\bar a^{e+p}_{\rm eff})_{0}-\alpha(\bar a^{n}_{\rm eff})_{0}$, with a small residual sensitivity to $(\bar a_{\rm eff}^{p+e+n})_{0}\equiv\alpha(\bar a^{e+p}_{\rm eff})_{0}+\alpha(\bar a^{n}_{\rm eff})_{0}$ proportional to deviations from the trend in $(A-Z)$ vs. $Z$.  The numerical factors multiplying these sum ($\propto A/m^{\rm At}$) and difference ($\propto (A-2Z)/m^{\rm At}$) coefficients are plotted for nuclides with lifetimes greater than one hour~\cite{nubase} in Figure~\ref{fig:aplot}.  Species which have been or may soon be used to test the EEP are explicitly indicated.  Also plotted are the coefficients for natural abundance SiO$_{2}$, since many modern gravimeters employ falling corner cubes made largely out of glass, and bulk Ti metal, as the best modern UFF tests compare it with $^{9}$Be~\cite{Schlamminger:2008}.

A UFF or matter-wave interferometer test which compares $^{1}$H with $^{3}$H or $^{4}$He would have the greatest intrinsic sensitivity to the $\alpha(\bar a^{w}_{\rm eff})_{0}$ coefficients.  If we restrict ourselves to heavier nuclides with equal  proton numbers, $^{6}$Li and $^{7}$Li are the clear favorites, with a suppression of only $.22$ on the difference term $(\bar a_{\rm eff}^{p+e-n})_{0}$, and $3.5\times 10^{-4}$ on the sum $(\bar a_{\rm eff}^{p+e+n})_{0}$.  Comparisons between $^{39}$K and $^{87}$Rb~\cite{rbvsk}, are nearly as sensitive, with a suppression factor of $0.19$ on the difference and $1.7\times 10^{-4}$ on the sum signals.  Comparisons between different stable isotopes of the same element become less sensitive with increased atomic weight.  A test comparing $^{6}$Li versus $^{133}$Cs or a $^{7}$Li versus any isotope of potassium would yield better sensitivity to the $(\bar a_{\rm eff}^{p+e+n})_{0}$ signal, with only a factor of $4.9\times 10^{-3}$ suppression.  The more recently analyzed $^{133}$Cs matter-wave redshift test~\cite{redshift} had a slightly greater sensitivity to $(\bar a_{\rm eff}^{p+e-n})_{0}$.

\section{Conclusion}

We have presented a quantitative analysis of the experimental signals for EEP violation in matter-wave interferometers in the context of the mSME, with a particular focus on anomalies that are difficult to constrain in non-gravitational experiments.  We find that it is unnecessary to exchange photons to carry out definitive tests of the gravitational redshift, as anomalous physics in the electromagnetic sector is either well constrained, or transferrable to other sectors by a judicious choice of coordinate chart.  We use the mSME to  quantitatively determine the relative sensitivities of existing and proposed experimental tests of the EEP~\cite{rbvsk}, illustrated in Figure~\ref{fig:aplot}.  This figure also reveals that tests employing one or more metastable nuclides can potentially offer greater sensitivity to these parameters than would otherwise be possible for stable isotopes with large $A$.  Matter-wave interferometers may be particularly well suited to carry out such tests, since the atomic source need not be isotopically pure, and particle decay on timescales longer than a single experimental shot (typically less than 10 s) will not affect the measured signal.

\appendix
\section{Equivalence to the Schr\"odinger equation}

From Sec. 2, it is clear that many effects in quantum mechanics are connected to the gravitational redshift and special relativistic time dilation. They can, therefore, be employed in testing general relativity. It is thus interesting to develop the above ideas into a more familiar form that is directly applicable to nonrelativistic quantum mechanics. Here, we will show that the interpretation of matter-wave interferometry as redshift tests is mathematically equivalent to the Schr\"odinger equation of an atom in a gravitational field. We shall follow the approach of Feynman \cite{Feynman1948}. This approach is, thus, not fundamentally new. However, there is pleasure in viewing familiar things from a new point of view. We start by using a post newtonian approximation
\be
S=\int mc^2 \sqrt{1+h_{\mu\nu}u^\mu u^\nu}d\lambda \approx  \int mc^2 \left(1+\tfrac 12 h_{00}+ h_{0j}\frac{u^j}{c}-\tfrac12(\delta_{jk}-h_{jk})\frac{u^j}{c}\frac{u^k}{c} \right)dt
\ee
where $u^j$ is the usual 3-velocity. We replaced the parameter $\lambda$ by the coordinate time $t$, which is possible at $O(c^{-3})$. We now compute the path integral for propagation over an infinitesimal distance between $\vec x_A$ and $\vec x_B$ over an infinitesimal time interval $\epsilon$, during which the integrand can be treated as constant. We denote $\vec \xi=\vec x_B-\vec x_A$. For an infinitesimal $\epsilon,$ $\vec v=\vec \xi/\epsilon$, so
\be
\psi(t+\epsilon,\vec x_A)=N\int d^3\xi\, \psi(t,\vec x_A-\vec \xi) \exp\left(i \frac{mc^2\eps}{\hbar}\left(1+\tfrac 12 h_{00}\right)\eps\right) \exp\left[-\frac{1}{2} A_{jk}\xi^j\xi^k + B_j\xi^j \right]
\ee
where $N$ is a normalization factor, $ A_{jk}\equiv -im (\delta_{jk}+h_{jk})/(\hbar \eps)$, and $B_j\equiv  im c h_{0j}/\hbar$. We can expand in powers of $\eps, \xi$:
\bea
\psi(t+\epsilon,\vec x_A)&=&N\int  \left(\psi-\xi^j \partial_j \psi+\tfrac12 \xi^j\xi^k \partial_j\partial_k \psi\right) \left[1+i \frac{mc^2\eps}{\hbar}(1+\tfrac12 h_{00})\right] e^{-\frac{1}{2} A_{jk}\xi^j\xi^k + B_j\xi^j}d^3\xi \nonumber
\eea
where $\psi\equiv \psi(t,\vec x_A)$. Computing the Gaussian integrals \cite{Zee}, we obtain
\bea
\psi(t+\epsilon,\vec x_A)&=&N\frac{(2\pi)^{3/2}}{\sqrt{\det A}} \left[\left(1+i \frac{mc^2\eps}{\hbar}(1+\tfrac12 h_{00})\right)\psi \right. \nonumber \\ && \left.
\quad\quad\quad\quad\quad-(\partial_j\psi)\frac{\partial}{\partial B_j}
+\frac 12 (\partial_j\partial_k\psi)\frac{\partial}{\partial B_j} \frac{\partial}{\partial B_k}  \right]  \exp\left(\frac12 B_jB_k(A^{-1})_{jk}\right),
\eea
where $\det A$ is the determinant of $A$ and $A^{-1}\approx -\frac{\hbar \eps} {im}(\delta_{jk}-h_{jk})$ is the inverse matrix. 
We determine the normalization factor $N= \sqrt{\det A}/(2\pi)^{3/2} \exp[\frac12 B_jB_k(A^{-1})_{jk}]$ from the fact that $\psi(t+\epsilon,\vec x_A)$ must approach $\psi(t,\vec x_A)$ for $\eps\rightarrow 0$ and carry out the derivatives with respect to $B_j$ and inserting $B_j$ and $(A^{-1})_{jk}$.
Working in post newtonian order 3, we can neglect $h_{jk}$ and terms proportional to $\eps^2$.  This leads to a Schr\"odinger equation
\be\label{Schrodinger}
i\hbar \frac{d}{dt} \psi =- mc^2\tfrac12 h_{00}\psi
- \frac{\hbar^2}{2m} \left(\vec \nabla-m\vec h \right)^2 \psi,
\ee
where we have substituted $\psi\rightarrow e^{-i \om_C t}\psi$. The 3-vector $\vec h$ is defined by $h_j\equiv (i c/\hbar)h_{0j}$. We neglected a term proportional to $h_{0j}h_{0j}$ and one proportional to $h_{0j,j}$.  From the path integral approach, the usual commutation relations can also be derived \cite{Feynman1948}. This shows that quantum mechanics is a description of waves oscillating at the Compton frequency that explore all possible paths through curved spacetime.


\end{document}